\documentclass{aa}  
\usepackage{graphicx}
\usepackage{txfonts}
\usepackage{natbib}
\newcommand{\psrtar}{PSR\,J0205$+$6449}

\newcommand{\gtap}{\mathrel{\hbox{\rlap{\lower.55ex \hbox {$\sim$}}
                   \kern-.3em \raise.4ex \hbox{$>$}}}}
\newcommand{\ltap}{\mathrel{\hbox{\rlap{\lower.55ex \hbox {$\sim$}}
                   \kern-.3em \raise.4ex \hbox{$<$}}}}

\begin{document}
   \title{Hard X-ray timing and spectral characteristics of the energetic pulsar \psrtar\ in supernova remnant 3C58}
   \titlerunning{Hard X-ray characteristics of the energetic pulsar \psrtar\ in 3C58}

   \subtitle{An RXTE PCA/HEXTE and XMM-Newton view on the 0.5-250 keV band}

   \author{L. Kuiper\inst{1}
          \and
          W. Hermsen\inst{1,2}
          \and
          J.O. Urama\inst{3}
          \and
          P.R. den Hartog\inst{1,4}
          \and
          A.G. Lyne\inst{5}
          \and
          B.W. Stappers\inst{5}
          }
   \offprints{L. Kuiper}

   \institute{SRON-Netherlands Institute for Space Research, Sorbonnelaan 2, 
              3584 CA, Utrecht, The Netherlands\\
              \email{L.M.Kuiper@sron.nl}
              \and
              Astronomical Institute ``Anton Pannekoek", University of 
              Amsterdam, PO Box 94249, 1090 GE, Amsterdam, The Netherlands\\
              \email{W.Hermsen@sron.nl}
              \and
              Dept. of Physics \& Astronomy, University of Nigeria, Nsukka\\
              \email{johnson@hartrao.ac.za}
              \and
              Stanford University HEPL/KIPAC Physics, 382 Via Pueblo Mall Stanford, 94305, USA\\
              \email{hartog@stanford.edu}
              \and
              Jodrell Bank Center for Astrophysics, School of Physics and Astronomy, The University of Manchester
              Manchester M13 9PL, UK
              \email{andrew.lyne@manchester.ac.uk; ben.stappers@manchester.ac.uk}
              }

   \date{Received 11 December 2009 / Accepted xxx 2010}

  \abstract
   {}
   {\psrtar\ is a young rotation-powered pulsar in SNR 3C 58. It is one of only three young ($<10,000$ year old) pulsars which are so far detected in the radio and the classical X-ray bands, as well as at hard X-rays above 20 keV and at high-energy ($>100$ MeV) $\gamma$-rays. The other two young pulsars are the Crab and PSR B1509-58. Our aim is to derive the timing and spectral characteristics of \psrtar\ over the broad X-ray band from $\sim$ 0.5 to $\sim$ 270 keV.}
   {We used all publicly available RXTE observations of \psrtar\ to first generate accurate ephemerides over the 
   period September 30, 2000 - March 18, 2006. Next, phase-folding procedures yielded pulse profiles using data from RXTE PCA and HEXTE, and XMM-Newton EPIC PN. All profiles have been phase aligned with a radio profile derived from the Jodrell Bank Observatory data, and the time-averaged timing and spectral characteristics of the pulsed X-ray emission have been derived.}
   {While our timing solutions are consistent with earlier results, our work shows sharper structures in the
    PCA X-ray profile. The X-ray pulse profile consists of two sharp pulses, separated in phase by $0.488 \pm 0.002$, 
    which can be described with 2 asymmetric Lorentzians, each with the rising wing steeper than the trailing wing, 
    and full-width-half-maximum $1.41 \pm 0.05$ ms and $2.35 \pm 0.22$ ms, respectively. 
    We find an indication for a flux increase by a factor $\sim 2$, about $3.5\sigma$ above the time-averaged value, for 
    the second, weaker pulse during a two-week interval, while its pulse shape did not change. The spectrum of the 
    pulsed X-ray emission is of non-thermal origin, exhibiting a power-law shape with photon index 
    $\Gamma = 1.03 \pm 0.02$ over the energy band $\sim$ 0.5 to $\sim$ 270 keV. In the energy band covered with the 
    PCA ($\sim 3-30$ keV) the spectra of the two pulses have the same photon index, namely, $1.04 \pm 0.03$ and 
    $1.10 \pm 0.08$, respectively. Comparisons of the detailed timing and spectral characteristics of \psrtar\ in the radio, hard X-ray and gamma-ray bands with those of the Crab pulsar, PSR B1509-58 and the middle-aged Vela pulsar do reveal more differences than similarities.}
   {}
   \maketitle
%

\section{Introduction}

\psrtar\ is a young rotation-powered pulsar of which the pulsations were first discovered
in X-rays in a 2002 Chandra X-ray Observatory (CXO) observation, reported by \citet{murray2002} 
together with a confirmation in an analysis of archival Rossi X-Ray Timing Explorer (RXTE) data. 
Subsequently, the weak radio signal was detected by \citet{camilo2002}. \psrtar\ is a young, 65-ms pulsar 
located in the center of supernova remnant/pulsar wind nebula (PWN)  3C 58. It is one of the most energetic pulsars in the Galaxy
with a spin-down luminosity $\dot{E}\sim{2.7}\times 10^{37}$erg s$^{-1}$, and characteristic age $\tau\sim{5.4}$ kyr. This characteristic age, estimated with the values of the period and period derivative,
puts in doubt the possible association with 3C 58, which coincides positionally with the historical 828 yr old supernova SN1191 \citep{stephenson2002}. However, an age of several thousand years for 3C 58, closer to the characteristic age of the pulsar, can be derived from the velocities of the radio expansion of the PWN \citep{bietenholz2006} and of optical knots \citep{fesen2008}.

Recently, \citet{liv2009} presented for \psrtar\ phase-coherent timing analyses using X-ray data from the Proportional Counter Array 
(PCA; 2-60 keV) aboard RXTE and radio data from the Jodrell Bank Observatory and the Green Bank Telescope (GBT), 
spanning together 6.4 yrs. This work revealed timing noise and two spin-up glitches. Furthermore, they presented detailed 
characteristics of the X-ray profile, which was detected up to $\sim$40 keV. Their X-ray profile template consisted of two Gaussian-shaped pulses, a narrow (full-width-half-maximum (FWHM) $\sim$ 1.6 ms), more intense pulse and a broader (FWHM $\sim$ 3.8 ms) weak pulse separated 0.505 in phase, the single radio pulse leading the main X-ray pulse by $ \phi = 0.10 \pm 0.01$. Earlier results from an analysis of part of the RXTE and GBT data were reported by \citet{ransom2004}. These authors also presented spectral fits over the energy band 3--16 keV for both pulses: the best fit power-law photon indices were hard, namely $\Gamma$ = $0.84_{-0.15}^{+0.06}$ for the main pulse and
$\Gamma$ = $1.0_{-0.3}^{+0.4}$ for the second (weaker) pulse.
 
Finally, high-energy $\gamma$-ray pulsations ($\geq$ 0.1 GeV) from \psrtar\ were discovered with the Large Area Telescope (LAT) aboard the Fermi Gamma-ray Space Telescope \citep{abdo2009a}, folding the $\gamma$-ray arrival times with the radio rotational ephemeris from, again, the GBT and Jodrell Bank. The $\gamma$-ray light curve for energies $\geq$ 0.1 GeV shows also two peaks with intensities differing by a factor $\sim$ 2, aligned with the X-ray peaks, However, the main X-ray pulse coincides in phase with the weakest $\gamma$-ray pulse which has the softest spectrum of the two at high-energy $\gamma$-rays. The total pulsed $\gamma$-ray spectrum exhibits a simple power-law shape with index $\Gamma \sim$ 2.1 and exponential cutoff at $\sim$ 3.0 GeV.

\psrtar\ is now one of only three young ($<10,000$ year old) pulsars which are detected in the classical X-ray band and at hard X-rays above 20 keV,  as well as at high-energy ($>0.1$ GeV) $\gamma$-ray energies, the others being the Crab pulsar and PSR B1509-58 (PSR J1513-5908). The Crab pulsar has been studied over the total high-energy band already in great detail \citep[see for a coherent high-energy picture from soft X-rays up to high-energy $\gamma$-rays][]{kuiper2001}, with even a detection of pulsed $\gamma$-rays above 25 GeV \citep{aliu2008}. 

The detection of pulsed emission above 100 MeV from PSR B1509-58 had to wait for the new generation of currently operational $\gamma$-ray telescopes \citep{pellizzoni2009}. However, these three young pulsars have very different timing and spectral characteristics. This makes it particularly interesting to determine the timing and spectral characteristics of \psrtar\ in more detail over the high-energy band of the electro-magnetic spectrum in order to compare these with those of Crab and PSR B1509-58 and for confrontation with theoretical predictions. In this work our aim is to extend the coverage in the hard X-ray band
to higher energies, exploiting the data of the High Energy X-ray Timing Experiment (HEXTE; 15-250 keV) aboard RXTE,
and to extend the energy window to lower energies by analysing data from XMM-Newton. We will present the results from 
our timing study exploiting only the multi-year PCA/RXTE monitoring data, which we performed in parallel to the work reported by \citet{liv2009}.
Our timing solutions are consistent with those of the latter authors, but our work revealed sharper structures in the PCA X-ray pulse profile. 
Furthermore, we derive the spectral characteristics over the total X-ray band. In the discussion we compare our findings  with  the characteristics of \psrtar\ reported in the radio band and at high-energy $\gamma$-rays, as well as with the timing and spectral characteristics of the Crab pulsar, PSR B1509-58 and the middle-aged Vela pulsar.

\section{Instruments and observations}

\begin{table}[t]
\caption{RXTE \psrtar\ observation summary}
\label{table_rxte_obs}
\centering
\begin{tabular}{c c c c c}
\hline\hline
Obs. id.    & Date begin & Date End   & MJD         & Exposure$^a$\\
            &            &            &             &   (ks)  \\
\hline
\hline
\vspace{-2mm}\\
20259       & 30-09-1997 & 30-09-1997 & 50721-50722 & 16.96   \\
\vspace{-2.5mm}\\
60130       & 17-08-2001 & 19-08-2001 & 52138-52141 & 80.35   \\
\vspace{-2.5mm}\\
70089       & 10-03-2002 & 23-04-2003 & 52343-52752 & 268.95  \\
\vspace{-2.5mm}\\
90080       & 28-02-2004 & 03-03-2005 & 53063-53432 & 243.23  \\
\vspace{-2.5mm}\\
91063       & 12-03-2005 & 18-03-2006 & 53441-53813 & 328.74  \\
\vspace{-2mm}\\
\hline
\vspace{-2mm}\\
\multicolumn{5}{l}{$^a$ Screened (GTI) exposure for PCA unit-2}\\ 
\end{tabular}
\end{table}

\subsection{RXTE}

In this study extensive use is made of data from monitoring observations
of \psrtar\/ with the two non-imaging X-ray instruments aboard RXTE, the Proportional Counter Array 
(PCA; 2-60 keV) and the High Energy X-ray Timing Experiment (HEXTE; 15-250 keV). The PCA 
\citep{jahoda96} consists of five collimated Xenon proportional 
counter units (PCUs) with a total effective area of $\sim 6500$ cm$^2$ over a $\sim 1\degr$ 
(FWHM) field of view. Each PCU has a front Propane anti-coincidence layer and three Xenon 
layers which provide the basic scientific data, and is sensitive to photons with energies in 
the range 2-60 keV. The energy resolution is about 18\% at 6 keV. All data used in this work have
been collected from observations in {\tt GoodXenon} or {\tt GoodXenonwithPropane} mode allowing 
high-time-resolution ($0.9\mu$s) studies in 256 spectral channels.

The HEXTE instrument \citep{rothschild98} consists of two independent detector 
clusters A and B, each containing four Na(Tl)/ CsI(Na) scintillation
detectors. The HEXTE detectors are mechanically collimated to a $\sim{1}\degr$ (FWHM) 
field of view and cover the 15-250 keV energy range with an energy resolution of 
$\sim$ 15\% at 60 keV. The collecting area is 1400 cm$^2$ taking into account the 
loss of the spectral capabilities of one of the detectors. The best time 
resolution of the tagged events is $7.6\mu$s. In its default operation mode the 
field of view of each cluster is switched on and off source to provide instantaneous 
background measurements.
Due to the co-alignment of HEXTE and the PCA, both instruments simultaneously observe the 
same field of view.

RXTE observed \psrtar\, for the first time on Sept. 30, 1997 (MJD 50721) for about 17 ks.
Data from this observation were used by \citet{murray2002} to confirm the pulsation discovered with Chandra.
A dedicated much deeper observation was performed in the period August 17-19, 2001 (MJD 52138-52141), yielding 
about 80 ks good exposure time. Next, a monitoring campaign started on March 10, 2003 (MJD 52343) which ended on
April 23, 2003 (MJD 52752). The total (good) exposure time for this period was about 269 ks. A second monitoring
round commenced on Feb. 28, 2004 and continued till March 18, 2006 (MJD 53063-53813) yielding a total (good) exposure
time of about 572 ks. A summary of all RXTE observations of \psrtar\ is given in Table \ref{table_rxte_obs}. The total
good-time exposure (after screening; see Sect. \ref{sect_tm_pca}) amounts 938.23 ks.

\subsection{XMM-Newton}

We searched the XMM-Newton observation database for observations of the field around \psrtar\ in which the EPIC-PN camera
\citep{struder2001} operated in Small-Window (SW) mode. This mode ($4\farcm4 \times 4\farcm4$ field of view) offers sufficient 
time resolution ($\sim 5.67$ ms) to sample the pulse-profile of \psrtar\ over the $\sim$ 0.3-12 keV range. We found two observations
(observation ids. 0004010101/0004010201) both performed on February 22, 2001 at $1\farcm2$ offset from \psrtar\ with durations of 
about 9.2 and 23.6 ks, respectively.

\begin{figure}[t]
  \centering
  \includegraphics[width=8.0cm,height=8.0cm]{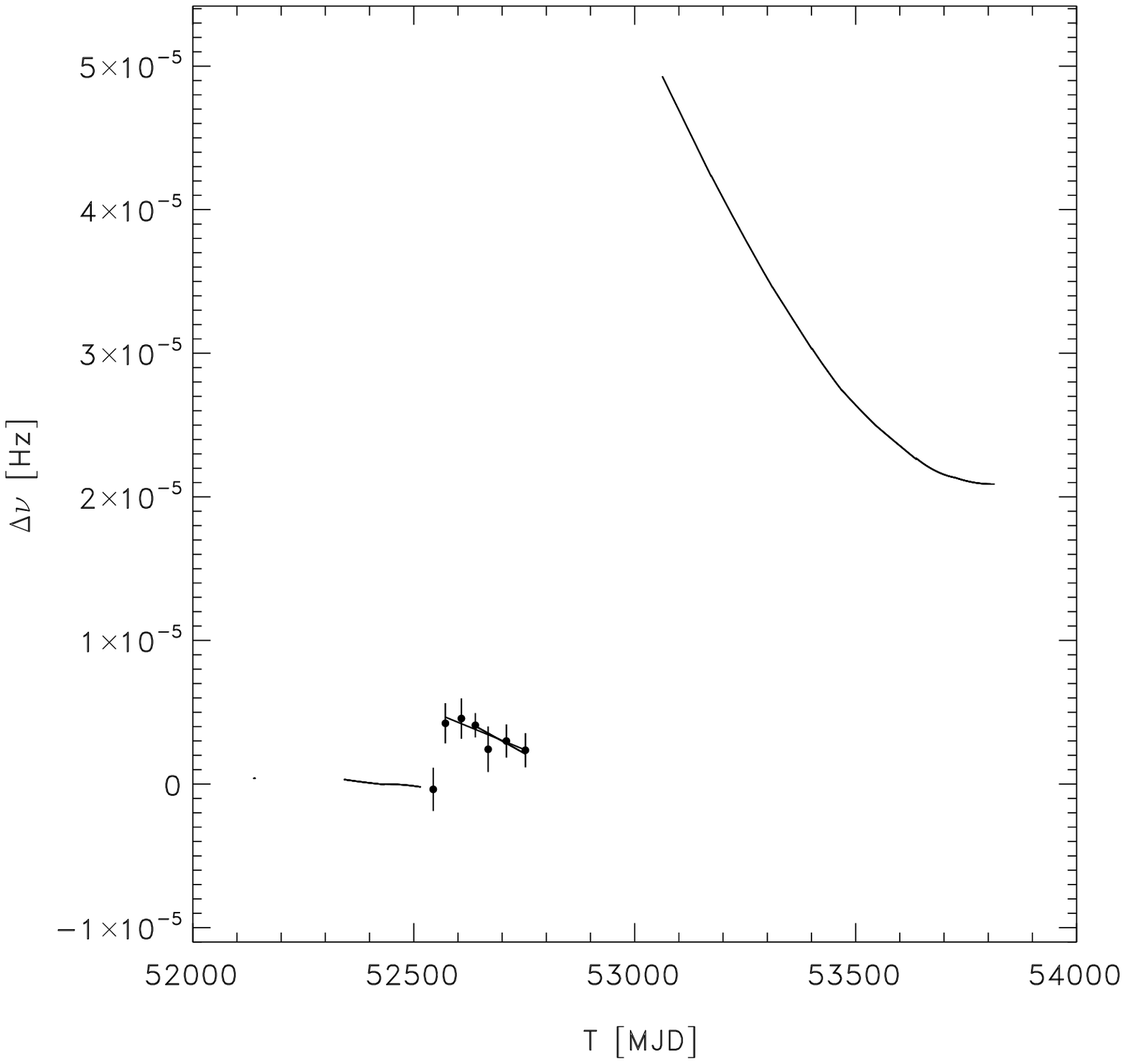}
  \caption{Evolution of the spin-frequency of \psrtar\ as derived from RXTE PCA data over the period MJD 52138-53813 with respect to the linear trend of
the phase coherent timing model of period MJD 52433-52515 (entry \#3 of Table \ref{eph_table}). As solid lines the entries \#1-12 of Table \ref{eph_table} are plotted, while for the period 52571-52752 also the linear fit to the incoherent frequency measurements is shown as solid line. Incoherent frequency measurements over the period MJD 52544-52752 are shown as data points. Note the presence of (at least) two timing glitches: one in the period 52515 to 52571 and a second stronger one between 52752 and 53063 (see text).
 }
  \label{freq_evol}
\end{figure}

\begin{table*}[t]
\caption{Phase-coherent ephemerides for \psrtar\ as derived from RXTE PCA (monitoring) data.}
\label{eph_table}
\begin{center}
\begin{tabular}{lccclllcr}
\hline
Entry &  Start &  End  &   t$_0$, Epoch   & \multicolumn{1}{c}{$\nu$}   & \multicolumn{1}{c}{$\dot\nu$}               & \multicolumn{1}{c}{$\ddot\nu$}                  & $\Phi_{0}^{3}$
  & Validity range\\
 \#   &  [MJD] & [MJD] &     [MJD,TDB]    & \multicolumn{1}{c}{[Hz]}    & \multicolumn{1}{c}{$\times 10^{-11}$ Hz/s}  & \multicolumn{1}{c}{$\times 10^{-21}$ Hz/s$^2$}  &           
  &  \multicolumn{1}{c}{(days)}   \\
\hline\hline
\\
0         & 50721 & 50722 & 50721.0     & 15.230153381(92) & -4.489  (fixed)        &   0.0 (fixed)              & 0.2931  & 2\\
\vspace{-2mm}\\
1         & 52138 & 52141 & 52138.0     & 15.224659060(16) & -4.489(16)             &   0.0 (fixed)              & 0.5389  & 4\\
\vspace{-2mm}\\
2         & 52343 & 52433 & 52343.0     & 15.223863557(3)  & -4.49605(13)           &   +1.96(38)                & 0.6540  & 91\\
3         & 52433 & 52515 & 52433.0     & 15.223514035(4)  & -4.49086(20)           &   -7.98(62)                & 0.7751  & 83\\
4$^1$     & 52639 & 52752 & 52639.0     & 15.2227187742(4) & -4.51063(1)            &   0.0 (fixed)              & 0.4147  & 114\\
\vspace{-2mm}\\
5         & 53063 & 53173 & 53063.0     & 15.2211188258(7) & -4.56381(2)            &   0.0 (fixed)              & 0.4814  & 111\\
6         & 53173 & 53312 & 53173.0     & 15.2206851395(13)& -4.55932(4)            &   +5.80(8)                 & 0.3170  & 140\\
7         & 53312 & 53401 & 53312.0     & 15.2201380229(35)& -4.54739(19)           &   +1.36(49)                & 0.3458  &  90\\
8         & 53401 & 53469 & 53401.0     & 15.2197884405(37)& -4.54444(25)           &   +12.0(9)                 & 0.3273  &  69\\
9         & 53469 & 53546 & 53469.0     & 15.2195216777(26)& -4.53041(15)           &   +7.18(47)                & 0.1970  &  78\\
10        & 53546 & 53637 & 53546.0     & 15.2192204400(23)& -4.52131(12)           &   +1.85(32)                & 0.6030  &  92\\
11        & 53637 & 53726 & 53637.0     & 15.2188650817(24)& -4.51951(12)           &   +28.5(3)                 & 0.1413  &  90\\
12        & 53726 & 53813 & 53726.0     & 15.2185183927(26)& -4.50300(14)           &   +16.8(4)                 & 0.2436  &  88\\
\vspace{-2mm}\\
\hline
\vspace{-2mm}\\
13$^2$    & 53726 & 53764 & 53745.0     & 15.2184445054(20)& -4.49904(15)           &   -10.3(61)                & 0.1147  &  39\\
14$^2$    & 53750 & 53814 & 53782.0     & 15.2183007045(13)& -4.49502(8)            &   +29.1(16)                & 0.6055  &  65\\
\hline
\multicolumn{9}{l}{$^1$ A glitch occured between MJD 52515 and 52571 \citep[see][for more information]{liv2009}. This entry describes}\\ 
\multicolumn{9}{l}{\,\, the last part of the glitch recovery period. Its validity is questionable given the low number of TOA's, namely 4, translating}\\
\multicolumn{9}{l}{\,\,  to 1 degree of freedom in the TOA fit procedure.}\\
\multicolumn{9}{l}{$^2$ Ephemeris from Jodrell Bank radio data}\\
\multicolumn{9}{l}{$^3$ $\Phi_{0}$ is the phase offset to be applied to obtain consistent radio-alignment (see Equation \ref{eq:phase} in Sect. \ref{sect_tm_xrc}) }\\
\end{tabular}
\end{center}
\end{table*}

\section{Timing}
\label{sect_tm}

\subsection{RXTE PCA timing analysis}
\label{sect_tm_pca}

The first step in the RXTE PCA data analysis was the screening of the data. We generated good-time intervals (GTI) for each PCU
separately, because the number of active PCU's at any instant was changing. 
Good time intervals have been determined for each PCU by including only time periods when the PCU in question is on, 
and during which the pointing direction is within $0\fdg 05$ from the target, the elevation angle above Earth's horizon 
is greater than $5\degr$, a time delay of 30 minutes since the peak of a South-Atlantic-Anomaly passage holds, and a low background level due to contaminating electrons is observed.
These good time intervals have subsequently been applied in the screening process to the data streams from each of 
the PCUs (e.g. see Table \ref{table_rxte_obs} for the resulting screened exposure of PCU-2 per observation run).

Next, we selected event data from {\em all} three Xenon layers of each PCU allowing us to better characterize the hard ($>10$ keV) 
X-ray properties of \psrtar.
The TT (Terrestial Time) arrival times of the selected events (for each sub-observation and for each PCU unit) have been converted to 
arrival times at the solar system barycenter (in TDB time scale) using 1) the JPL DE200 solar system ephemeris, 2) 
the instantaneous spacecraft position and 3) the sub-arcsecond celestial position of \psrtar.
The position used is:  $(\alpha,\delta)=(02^{\hbox{\scriptsize h}}05^{\hbox{\scriptsize m}}37\fs92,+64\degr49\arcmin42\,\farcs8)$ 
for epoch J2000 \citep{slane2002}, which corresponds to (l,b)=(130.71931,3.08456) in Galactic coordinates. 

\subsection{Timing solutions: ephemerides}
\label{sect_tm_sol}
We generated pulsar timing models (ephemerides) specifying the rotation behaviour of the pulsar over a certain time stretch.
The pulse frequency and its first two time derivatives $(\nu,\dot\nu,\ddot\nu)$ were determined from PCA X-ray data solely\footnote{X-ray timing data are not hampered by time-variable dispersion measure (DM) variations as is the case for the radio data, and therefore the remaining scatter in the obtained timing solutions is less. This has been verified for the Crab pulsar.}, demanding
a maximum RMS value of only 0.01 period in the time-of-arrival (TOA) analysis. This requirement resulted in 13 timing models with 
validity intervals of typically 100 days. The ephemerides are listed in Table \ref{eph_table}. In the TOA analysis we followed the steps 
outlined in Section 4 of \cite{kuiper2009}, in this case, however, we made a high-statistics correlation template (showing clearly the two
X-ray pulses) from the 80 ks observation during run 60130. Our models are fully consistent with those derived by \citet{liv2009}, who used a combination of X-ray (RXTE PCA) and radio (GBT and JBO) data. Also, we found evidence for the presence of two timing glitches using solely X-ray data, one occuring somewhere between MJD 52515 and 52571 and a much stronger one occuring in the RXTE monitoring gap between MJD 52752 and 53063 \citep[see][for more details on these glitches which they report to have fractional magnitudes ${\Delta}{\nu}/\nu \sim 3.4 \times 10^{-7}$ and ${\Delta}{\nu}/\nu \sim 3.8 \times 10^{-6}$, respectively]{liv2009}. 
The frequency evolution history over the RXTE observation time stretch MJD 52138-53813 is shown in Fig. \ref{freq_evol}.

The main difference between our work and that performed by \citet{liv2009} is that we chose for an accurate (RMS $<0.01$) description of the rotation behaviour of the pulsar with at most 3 timing parameters over a limited time stretch in stead of using many more timing parameters 
over a much wider time interval. In the latter approach the need for the (unphysical) higher order timing parameters reflects the presence of (strong) timing noise.

\subsection{X-ray/radio pulse profile phase alignment}
\label{sect_tm_xra}
The Jodrell Bank observatory (JBO) made observations at a radio frequency of 1.4 GHz from MJD 53725 to 54666, and therefore overlaps
for about 89 days with the second RXTE monitoring cycle in the period MJD 53725 to 53813. For two time segments in this interval, MJD 53726-53764
(number of TOAs, 28) and MJD 53750-53814 (number of TOAs, 29), accurate (RMS $<0.01$) timing models are constructed with 3 timing parameters (see also
Table \ref{eph_table}).
The 1.4 GHz single-pulse radio profile (in 400 bins) is shown in Fig. \ref{xrg_collage}a with a fiducial point (defining radio-phase 0.0) 
corresponding to the centre of gravity of the single pulse (just before the pulse maximum). Folding the barycentered X-ray time tags from period MJD 53726 to 53813 upon these radio-ephemerides put the main X-ray pulse (pulse-1) at phase $0.089\pm0.001$ (statistical error only), consistent with the value quoted for the JBO-PCA offset in \citet{liv2009}, namely $0.085\pm 0.010$.
Next, we determine through correlation analysis the phase shifts to be applied to the X-ray pulse profiles from the data periods of entries 0-12 of Table \ref{eph_table} to align these to the radio-aligned X-ray profile of period MJD 53726-53813.
These shifts ($\Phi_0$) are given in Table \ref{eph_table}.

\subsection{Combined X-ray event matrix from PCA observations}
\label{sect_tm_xrc}
\begin{figure}[t]
  \centering
  \includegraphics[width=8.5cm,height=8.0cm,bb=60 170 555 660,clip=]{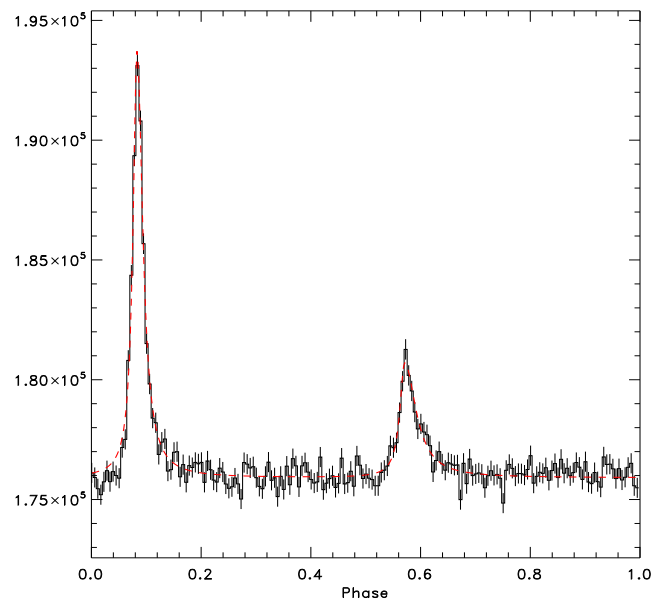}
  \caption{High-statistics radio-aligned PCA pulse phase distribution in 180 bins for PHA range 5-44 ($\sim$2-20 keV) combining all available radio-aligned pulse phase distributions from different data segments. {\bf Error bars represent $1\sigma$ uncertainties.} The best fit ($\chi^2=199.47$ for 171 degrees of freedom) model composed of two asymmetric Lorentzians plus background is superposed {\bf as dashed red line}. The rising wings of the pulses are steeper than the trailing wings.}
  \label{xray_pulseprofile}
\end{figure}

Barycentered PCA X-ray event times are finally folded upon an appropriate timing model composed of $\nu,\dot\nu,\ddot\nu$ and 
the epoch $t_0$, as shown in  Table \ref{eph_table}. Proper X-ray/radio phase alignment, $\Phi(t)$, is obtained by subtracting $\Phi_0$ as shown in the following formula: \begin{equation}\Phi(t)=\nu\cdot (t-t_0) + \frac{1}{2}\dot\nu\cdot (t-t_0)^2+\frac{1}{6}\ddot\nu\cdot (t-t_0)^3 - \Phi_0 \label{eq:phase}\end{equation}
Combining the radio-aligned phase information for all PCA-data covered with a proper ephemeris (see Table \ref{eph_table}) yielded an event matrix 
N($\Phi$,E) of (180$\times$256) elements. For this purpose we binned the pulse-phase interval [0,1] into 180 phase bins for all 256 PCA PHA channels.
From this matrix a high-statistics pulse-phase distribution was extracted for the PHA channels 5-44 ($\sim$ 2-20 keV). This distribution is shown in Fig. \ref{xray_pulseprofile}.

\subsection{X-ray pulse profile characterization}
\label{sect_tm_xrpc}

Initially, we fitted, analogous to \citet{liv2009}, our high-statistics RXTE PCA pulse profile shown in Fig. \ref{xray_pulseprofile} with a model consisting of 2 Gaussians, each with free scale, width and position, plus background. However, this model rendered a poor/unacceptable fit ($\chi^2=265.48$ for 180 - 7 degrees of freedom). Next, we tried a double symmetric Lorentzian model plus background in order to give more weight to the wings of the pulses. This model provided
a better description of the measured pulse-phase distribution ($\chi^2=239.13$ for 180 - 7 degrees of freedom), but is still poor. Finally, we
abandoned the description in terms of symmetric functions and used a combination of 2 asymmetric Lorentzians plus background. This model (9 
free parameters) is specified below:

\begin{table}[t]
\caption{X-ray pulse profile characterization of \psrtar\ from a fit involving two asymmetric Lorentzians plus background}
\label{best_fit_table}
\begin{center}
\begin{tabular}{llcl}
\hline
Parameter &  Value &     &   $1\sigma$-error\\
\hline\hline\\
\multicolumn{4}{c}{Pulse-1}\\
\vspace{-2mm}\\
$\Phi_1^{a}$       & 0.0831  & $\pm$ & 0.0004\\
$\Gamma_{1l}$ & 0.0175  & $\pm$ & 0.0009 \\
$\Gamma_{1r}$ & 0.0252  & $\pm$ & 0.0011 \\
\vspace{-2mm}\\
\multicolumn{4}{c}{Pulse-2}\\
\vspace{-2mm}\\
$\Phi_2$       & 0.5709  & $\pm$ & 0.0015\\
$\Gamma_{2l}$ & 0.0214  & $\pm$ & 0.0036 \\
$\Gamma_{2r}$ & 0.0502  & $\pm$ & 0.0056 \\
\vspace{-2mm}\\
\multicolumn{4}{c}{Derived quantities}\\
\vspace{-2mm}\\
$\Phi_2-\Phi_1$ & 0.488  & $\pm$ & 0.002\\
$N_1/N_2$       & 3.72   & $\pm$ & 0.23\\
$I_1^{b}$       & 0.688  & $\pm$ & 0.011\\
$I_2$           & 0.312  & $\pm$ & 0.014\\
$R=I_1/I_2$     & 2.2    & $\pm$ & 0.1\\
$\Gamma_1=(\Gamma_{1l}+\Gamma_{1r})/2$      & 0.0214 & $\pm$ & 0.0007\\
                                              & 1.41   & $\pm$ & 0.05 ms\\
$\Gamma_2=(\Gamma_{2l}+\Gamma_{2r})/2$      & 0.0358 & $\pm$ & 0.0034\\
                                              & 2.35   & $\pm$ & 0.22 ms\\
\vspace{-2mm}\\
\hline\hline\\
\multicolumn{4}{l}{$^a$ Statistical error only, the systematic error is of the }\\
\multicolumn{4}{l}{\ \  order of 0.01 \citep[see][]{liv2009} }\\
\multicolumn{4}{l}{$^b$ Relative contribution of the integrated flux in pulse-1}\\
\multicolumn{4}{l}{\, \,to the total pulsed flux}\\
\end{tabular}
\end{center}
\end{table}

\begin{eqnarray}
N(\phi;B,\vec{p_1},\vec{p_2})=B+{\cal{N}}_1(\phi;\vec{p_1})+{\cal{N}}_2(\phi;\vec{p_2})
\end{eqnarray}


In this formula $\vec{p_1}$ represents the 4 model parameters, $(N_1,\phi_1,\Gamma_{1l},\Gamma_{1r})$, of the first asymmetric Lorentzian,
$\vec{p_2}$ the  equivalent parameters, $(N_2,\phi_2,\Gamma_{2l},\Gamma_{2r})$, describing the second asymmetric Lorentzian and $B$ is the
value of the background level. The first asymmetric Lorentzian is described by the following expression:

\begin{eqnarray}
{\cal{N}}_1(\phi;\vec{p_1})= \left\{ \begin{array}{lcr}
                      \frac{N_1}{\left((\phi-\phi_1)/(\Gamma_{1l}/2)\right)^2+1} & \ \  & \phi \le \phi_1\\
                                                                                     & \ \   &              \\
                      \frac{N_1}{\left((\phi-\phi_1)/(\Gamma_{1r}/2)\right)^2+1} & \ \  & \phi > \phi_1\\
                      \end{array}
               \right.
\end{eqnarray}

In this description $N_1$ is the maximum value of pulse-1 reached at $\phi_1$, the location of the maximum of pulse-1, $\Gamma_{1l}/2$ is the width of the left wing of the pulse-1, and finally $\Gamma_{1r}/2$ is the width of the right wing of the pulse-1.
A similar expression and equivalent definitions hold for the second asymmetric Lorentzian.

This composite model provided an excellent fit, $\chi^2=199.47$ for 171 degrees of freedom, with best-fit parameters and their $1\sigma$ error estimates
listed in Table \ref{best_fit_table} (see also the best-fit model superposed on the data in Fig. \ref{xray_pulseprofile}). A description in terms of two asymmetric Lorentzians plus background provides a $7.7\sigma$ improvement over the
two-Gaussians-plus-background model and a $6\sigma$ improvement over the two-Lorentzians-plus-background model, taking into account the two (=9-7) additional degrees of freedom in both cases. Therefore, our analysis does not support the assumption made by \citet{liv2009} of an underlying double Gaussian shape for the X-ray profile. We find the X-ray pulses to be sharper, especially for pulse-2. For both pulses the rising wings are significantly steeper than the trailing wings.

\begin{figure}[t]
 \centering
 \includegraphics[width=7.cm,height=13cm,bb=170 140 420 665,clip=]{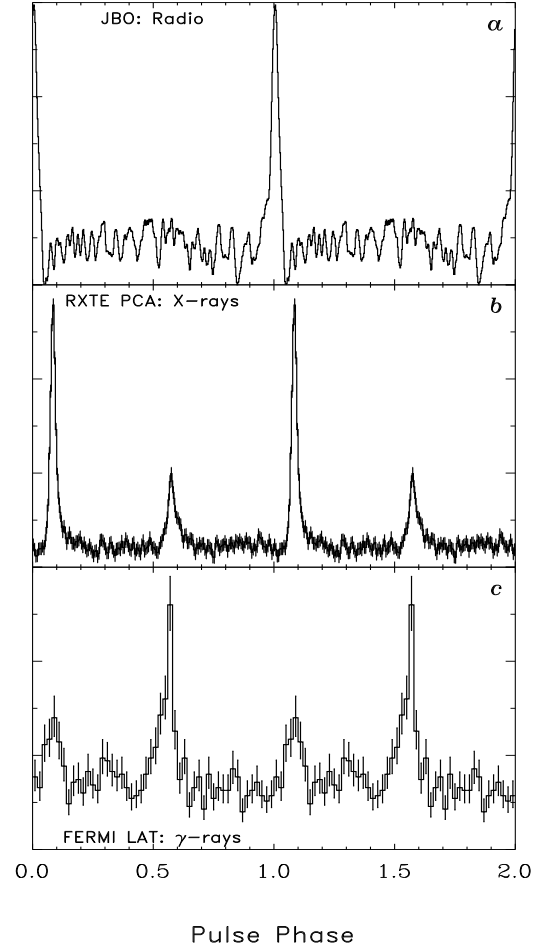}
 \caption{A comparison in absolute phase of the JBO radio (1.4 GHz), RXTE-PCA ($\sim$2-20 keV) and Fermi LAT ($>100$ MeV) pulse profiles.
          Note the change in the relative contributions of P1 and P2 in the X-ray and $\gamma$-ray windows.}
  \label{xrg_collage}
\end{figure}

The X-ray peak separation $\phi_2-\phi_1$ derived in this work is $0.488(2)$, significantly smaller than the value estimated by \citet{liv2009}, but consistent with the separation of $0.49 \pm 0.01 \pm 0.01$\footnote{The first error specifies the statistical error and the second the systematical error \citep[see][for more details]{abdo2009a}} measured at high-energy $\gamma$-rays by \citet{abdo2009a} using Fermi LAT $>100$ MeV data. The comparison of the shapes and absolute phases of the JBO radio, our RXTE-PCA X-ray and the Fermi-LAT profiles is shown in Fig. \ref{xrg_collage}. The main X-ray pulse (P1) appears to be the sharpest pulse in this comparison.

\subsection{X-ray pulse profile variability}
\label{sect_tm_xrpv}

We investigated the stability of the X-ray pulse-shape as a function of time. Therefore, we fitted the measured X-ray pulse-phase distribution 
(PHA range [4,27] $\sim$ 2-11 keV) for 15 time periods in terms of a constant background and the shapes of pulse-1 and pulse-2, separately. 
The times of these data points correspond to those of the X-ray timing models shown in Table \ref{eph_table} (entries 0 to 11; 12 points), 
augmented with two measurements during the last RXTE monitoring period, MJD 53726-53813, covering entry \#12 and finally, with a data point 
covering the post-glitch-1 period MJD 52544-52607, yielding eventually 15 independent measurements. The splitting of period MJD 53726-53813 into
the intervals MJD 53736-53749 (2 RXTE sub-observations) and MJD 53760-53813 (5 sub-observations) was driven by the detection of a ``timing
anomaly'' in the former interval during the phase-coherent timing analysis. At a later stage of this work it turned out that this ``anomaly'' was caused by incorrect RXTE clock corrections just after the introduction of a leap second on 2006, January 1.

\begin{figure}[t]
  \centering
  \hspace{-0.55cm}\includegraphics[width=5.5cm,height=9.45cm,angle=90]{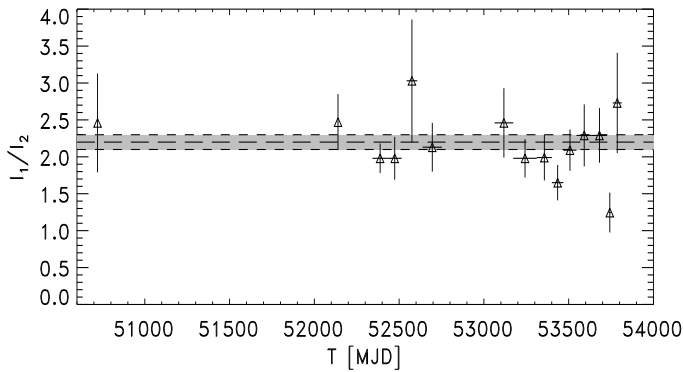}
  \caption{The ratio of the integrated flux in pulse-1 over pulse-2 as a function of time for the PCA energy band $\sim$ 2-11 keV.
           One data point at time interval MJD 53736-53749 (i.e. the ``anomaly'' period) deviates $\sim 3.5\sigma$ (single trial) from the 
           time-averaged value of $2.2\pm0.1$.}
  \label{i1_i2_ratio}
\end{figure}

\begin{figure}[t]
  \centering
  \includegraphics[width=7.5cm,height=7.5cm,bb=75 191 515 630,clip=]{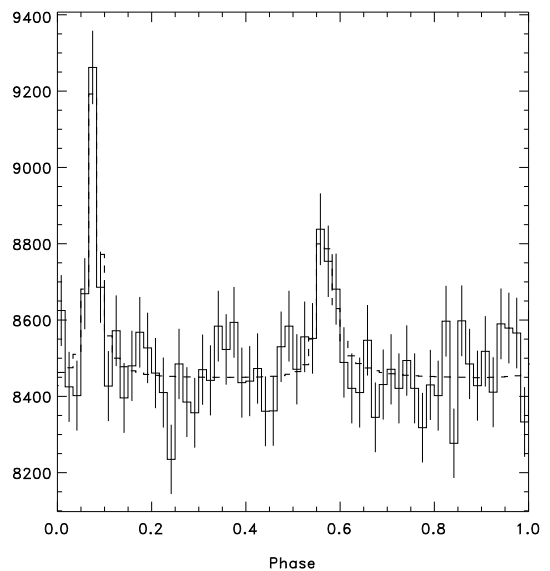}
  \caption{The pulse profile of \psrtar\ (60 bins) in the $\sim$2-11 keV band during the ``anomaly'' interval (MJD 53736-53749; 2 RXTE sub-observations).
           Strongly enhanced P2 emission is detected. The best fit model, composed of a background plus two asymmetric Lorentzians of the same shapes as shown in Fig. 2, is superposed as dashed line.}
  \label{anomaly_profile}
\end{figure}

The profile fitting procedure yields the flux ratio $R=I_1/I_2$ (see for the definition Table \ref{best_fit_table}) for each time interval. 
The results, $R(t)$ vs. $t$, are shown in Fig. \ref{i1_i2_ratio} with superposed as long-dashed line the P1/P2-flux ratio from the time-averaged high-statistics profile ($R=2.2\pm0.1$; see Table \ref{best_fit_table}) along with its $1$-$\sigma$ error region (shaded area).
One data point, corresponding to the ``anomaly'' period, deviates $\sim3.5\sigma$ from the time-averaged value. Taking into account the number of trials (15) its significance reduces to $2.7\sigma$, still indicating an interesting hint for variability. The pulse-phase distribution during the ``anomaly'' period is shown in Fig. \ref{anomaly_profile}. In this figure we also superposed as dotted line the best fit model in
which the shapes (two asymmetric Lorentzians) for each of the two pulses are identical to those derived for the time-averaged profile in Fig. \ref{xray_pulseprofile} and detailed in Table \ref{best_fit_table}.

Compared to the other measurements, where P2 is sometimes hardly visible, we see strongly enhanced P2 emission during this period. From spectral analysis of the P1 and P2 emissions during the ``anomaly'' period it turns out that the P1 flux is comparable to its time-averaged value contrary to the P2 flux, which shows a clear enhancement by almost a factor 2.

This leads to the conclusion that we see an interesting indication for flux variability for P2 without a change of its pulse shape.
Finally, we checked the JBO radio profile assembled during the ``anomaly'' period for possible morphology changes e.g. the appearance of a new
feature, but we found none.

\subsection{RXTE HEXTE timing analysis}
\label{sect_tm_hexte}
HEXTE operated in its default rocking mode during the observations listed in Table \ref{table_rxte_obs}, allowing the 
collection of real-time background data from two independent positions $\pm 1\fdg 5$ to either side of the 
on-source position. For the timing analysis we selected only the on-source data from both clusters.
Good-time intervals have been determined using similar screening filters as used in the case of the PCA. 
The selected on-source HEXTE event times have subsequently been barycentered and folded upon the ephemerides 
listed in Table \ref{eph_table} taking into account proper radio-phase referencing.
Thus, we obtained time-averaged HEXTE pulse phase distributions in 256 spectral channels 
(15 - 250 keV) for the combination of observations listed in Table \ref{table_rxte_obs}.
The total dead-time corrected exposure time collected for clusters A and B amounts, 400.6 ks and 426.3 ks, respectively.
Pulse profiles for the bands\footnote{We ignored spectral data from the band 28-33.1 keV because of the presence of a huge 
background line originating from the activation of Iodine.}, 14.7-28 and 33.1-132.6 keV, are shown in panels c and d of Fig. 
\ref{xmm_hexte_collage}. Fitting a model, comprising the (asymmetric) Lorentzians shapes of Pulses 1 and 2 and a flat 
background, to these phase distributions yielded detection significances of $7.8\sigma$ and $3.8\sigma$ for the 14.7-28 and 
33.1-132.6 keV bands, respectively ($2.9\sigma$ for the band, 64.1-132.6 keV). 
Therefore, pulsed emission of \psrtar\ has been detected up to $\sim 132$ keV, well above the sensitivity band of the PCA.

\subsection{XMM-Newton timing analysis}
\label{sect_tm_xmm}
\begin{figure}[t]
 \centering
 \includegraphics[width=7.5cm,height=13cm,bb=170 138 420 653,clip=]{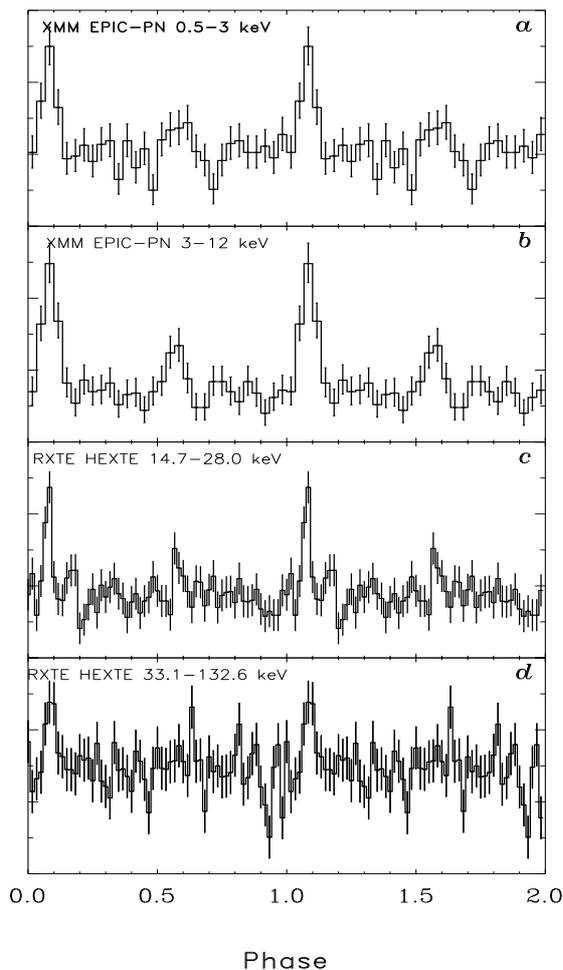}
 \caption{XMM-Newton EPIC-PN pulse profiles (30 bins) of \psrtar\ for the 0.5-3 and 3-12 keV energy bands (panels a,b).
          Panels c and d show the RXTE HEXTE profiles (60 bins) for the 14.7-28 and 33.1-132.8 keV energy bands.
          Significant pulsed emission is detected up to $\sim 130$ keV and down to $\sim 0.95$ keV.}
  \label{xmm_hexte_collage}
\end{figure}

The XMM EPIC-PN data were screened for solar (soft proton) flares by creating a light curve for events with energies in excess of 10 keV.
From the resulting count rate distribution, assumed to be Gaussian in absence of any flares, we could identify periods during which
the rate exceeds its mean value plus three times the width of the distribution. These periods are ignored in subsequent analysis. 
Next, we selected events from a sufficiently large circular region centered on \psrtar\ with a radius of $60\arcsec$ to ensure that all pulsar counts are included and barycentered the event times of these events.
Because the XMM-Newton observations have been performed before the 80 ks RXTE observation (60130) no valid ephemeris was available
for the XMM data period. Therefore, we performed a limited periodicity search around the predicted frequency value based on entry-1 of
Table \ref{eph_table}. We found a $\sim 10\sigma$ signal right at the predicted frequency using only events with energies between 4 and
12 keV. The folded pulse profile was compatible with the high-statistics PCA profile (see Fig. \ref{xray_pulseprofile}) taking into 
account the blurring of the pulse profile due to the limited time resolution of 5.67 ms (=0.086 in phase units). For subsequent 
studies we used an extraction radius of $15\arcsec$ because the signal-to-noise ratio peaks at that value. 
We created pulse-phase distributions, combining both adjacent XMM observations, in 30 bins (i.e. oversampled by a factor of $\sim 3$) for 585 energy intervals over the 0.3 to 12 keV range, each 0.02 keV wide.
The (radio-aligned) EPIC-PN pulse profiles for the 0.5-3 and 3-12 keV bands are shown in the panels a and b of Fig. \ref{xmm_hexte_collage} and have significances of $5.8$ and $9.7\sigma$ \citep[adopting $Z_7^2$-test;][]{buccheri1983}, respectively. Pulsed emission has been detected down to $\sim 0.95$ keV. 

\section{Pulsed spectra from RXTE PCA/HEXTE and XMM-Newton data: Total pulsed, P1 and P2}
\label{sect_tm_spectra}
From the pulse-phase distributions $N(\Phi,E)$, derived for RXTE PCA, HEXTE and XMM-Newton EPIC PN, we
extracted pulsed excess counts by fitting the following model function (analoguous to our X-ray pulse profile variability study shown in Sect. \ref{sect_tm_xrpv}) to the measured pulse-phase distribution, ${\cal{N}}(\Phi)$,
in various user-selected energy bands:
\begin{equation}
{\cal{N}}(\Phi)=b+c_1\times T_1(\Phi)+c_2\times T_2(\Phi) 
\end{equation}
In this formula $b$ represents the (constant) unpulsed/DC level, $c_1$ and $c_2$ the scales of the two asymmetric Lorentzian templates, $T_1$ and
$T_2$ (both normalized to 1), respectively (see Sect. \ref{sect_tm_xrpc}). This model provided statistically good fits
to all PCA and HEXTE profiles. Good fits could be derived for the EPIC-PN profiles after convolving the Lorentzian templates
with the poorer time resolution.

For each instrument the pulsed excess counts in the various energy bands for the first (P1) and the second pulse (P2) and the sum (=total pulsed, TP) can be translated to photon fluxes provided that proper energy response matrices are used.

In the case of the PCA we constructed time-averaged energy response matrices for each PCU separately taking into account the different (screened) exposure times of the involved PCU's during the time period of interest. For this purpose we used the {\it ftools version 6.4} programs {\it pcarsp} and {\it addrmf}. To convert PHA channels to measured energy values, $E_{\hbox{\scriptsize PHA}}$, for PCU combined/stacked products we also generated a weighted PCU-combined energy response matrix.

For HEXTE we employed cluster A and B energy-response matrices separately, taking into account the different screened on-source exposure times and the reduction in efficiency in case of off-axis observations. The on-source exposure times for both clusters have been corrected for considerable dead-time effects.

Finally, we created energy response files (effective area ({\em arf}) and energy redistribution matrix ({\em rmf})) for the EPIC PN operating in small window mode taking into account the reduction in effective area given the $15\arcsec$ source extraction radius used. For this purpose we employed the XMM SAS (vrs. 7.1.0) software tools {\em arfgen 1.73.3}
and {\em rmfgen 1.55.1}.

We assume simple power-law models in the form, $F_{\gamma}= K\cdot (E_{\gamma}/E_0)^{-\Gamma}$ with $\Gamma$ the photon-index and $K$ the normalization in ph/cm$^2$s keV at the pivot energy $E_0$, for the underlying photon spectra of P1, P2 and its sum TP. We fixed the absorbing interstellar Hydrogen column N$_{\hbox{\scriptsize H}}$ to $3.4\times 10^{21}$ cm$^{-2}$ \citep[see the ``PL-model for neutron star'' entry in Table 2 of][]{slane2004}.
These models have been fitted in a forward folding procedure using the appropriate response matrices to obtain the optimum spectral parameters, $K$ and $\Gamma$, and the reconstructed spectral flux points from the observed pulsed count rates. We verified that the measured high-statistics RXTE-PCA spectrum, as well as the EPIC-PN spectrum and the total spectrum including also HEXTE data are fully consistent with this non-thermal
simple power-law model. There is in the pulsed X-ray spectrum above $\sim$ 0.5 keV no indication for a thermal black-body component, a conclusion also reached for the total emission from the compact source by \citet{slane2002} and \citet{slane2004}, who reported a power-law spectral index of $\sim$ 1.7. We note that \citet{kargaltsev2008} in their Table I erroneously mark this pulsar to have a black-body component. Furthermore, in this work we only show the unabsorbed spectra i.e. the interstellar absorption has been modeled out.

In Table \ref{spectral_fit} the best fit values are listed of the spectral parameters for the total pulsed emission TP,
and emissions of P1 and P2 using PCA data only, and for TP using the EPIC PN, PCA and HEXTE combination over the extended energy band 0.56 to 267.5 keV. All spectra have a consistent shape with index $\sim 1.03$. We note for comparison, that for energies $>100$ MeV the Fermi LAT measures for TP a much softer spectrum with spectral index 2.1 with cutoff at $\sim$ 3 GeV \citep{abdo2009a}, and that P2 exhibits at high-energy $\gamma$-rays a significantly harder spectrum than P1.

The photon spectrum ($\nu F_{\nu}$ representation) over the 0.56 - 267.5 keV energy band of the total pulsed emission combining XMM-Newton EPIC-PN, RXTE-PCA and HEXTE data, as derived in this work, is shown in Fig. \ref{he_spectrum} in a much wider energy frame (0.1 keV-10 GeV) by including the best fit and its uncertainty range measured by Fermi for energies $>100 $ MeV \citep{abdo2009a}. The luminosity of the pulsed emission of \psrtar\ apparently reaches a maximum in the MeV band. For comparison are also shown the total pulsed emission spectra of the Crab, PSR B1509-58 as well as the ``middle-aged" Vela pulsar.

{\tabcolsep=1.5mm
\begin{table}[t]
\caption{Best fit values for the photon flux spectra of the total pulsed emission (TP), the first (P1) and second (P2) pulse emissions of \psrtar\ assuming a power-law
model of the form $F_{\gamma}= K\cdot (E_{\gamma}/E_0)^{-\Gamma}$.}
\label{spectral_fit}
\begin{center}
\begin{tabular}{lccc}
\hline
Parameter &  TP &  P1   &  P2\\
\hline\hline\\
\multicolumn{4}{c}{PCA\ \ (2.5-54.0 keV) }\\
\vspace{-2mm}\\
$K$ ($10^{-6}$ ph/cm$^2$s keV)       & $2.93 \pm 0.05$ & $2.03 \pm 0.03$ & $0.90 \pm 0.04$\\
$\Gamma$                             & $1.06 \pm 0.03$ & $1.04 \pm 0.03$ & $1.10 \pm 0.08$\\
$E_0$ (keV)                          & $8.34$          & $8.34$          & $8.34$         \\
$F_{2-30}$ ($10^{-13}$ erg/cm$^2$s)  & $10.67\pm 0.16$ & $7.43\pm 0.11$  & $3.24\pm 0.13$ \\
\vspace{-2mm}\\
\multicolumn{4}{c}{XMM/PCA/HEXTE\ \  (0.56-267.5 keV) }\\
\vspace{-2mm}\\
$K$ ($10^{-6}$ ph/cm$^2$s keV)          & $2.85 \pm 0.04$ &                 &                \\
$\Gamma$                                & $1.03 \pm 0.02$ &                 &                \\
$E_0$ (keV)                             & $8.49$          &                 &                \\
$F_{0.5-150}$ ($10^{-12}$ erg/cm$^2$s)  & $5.48\pm 0.28$  &                 &                \\
\vspace{-2mm}\\
\hline\hline\\
\end{tabular}
\end{center}
\end{table}
}

\section{Summary}
\label{sect_summary}
In this paper we derived for the young rotation-powered pulsar \psrtar\ the timing and spectral characteristics over the broad X-ray band from $\sim$ 0.5 to $\sim$ 270 keV, using data from the RXTE PCA and HEXTE, and XMM-Newton EPIC PN. These X-ray characteristics complement our knowledge about this pulsar in the radio domain and the high-energy $\gamma$-ray band for energies above 100 MeV.

-- Our phase-coherent ephemerides (see Table \ref{eph_table}) are consistent with those derived by \citet{liv2009} with the main difference that we used solely X-ray data (from the RXTE PCA) and fitted at most three timing parameters $(\nu,\dot\nu,\ddot\nu)$ over more limited time intervals. 

-- The X-ray pulse profile consists of two sharp pulses which can be described with 2 asymmetric Lorentzians, each with the rising wing steeper than the trailing wing, and full-width-half-maximum 1.41 $\pm$ 0.05 ms and 2.35 $\pm$ 0.22 ms, respectively. These profiles are sharper than reported by \citet{liv2009}.

-- The first X-ray pulse lags the single radio pulse in phase by 0.089 $\pm$ 0.001 (statistical error); the phase separation between the two X-ray pulses amounts 0.488 $\pm$ 0.002, fully consistent with the value 0.49 $\pm$ 0.01 $\pm$ 0.01  (statistical and systematic errors) reported for high-energy $\gamma$-rays above 100 MeV \citep{abdo2009a}.

-- We find an indication for a flux increase by a factor $\sim$ 2, $\sim3.5\sigma$ above the time-averaged value, for the second, weaker pulse during a two-week time interval, while its pulse shape did not change. During this time window, the morphology of the JBO radio profile of \psrtar\ did not change, notably, there was no indication for a second pulse.

-- We detected the pulsed signal significantly for the first time down to $\sim$ 0.95 keV with XMM-Newton EPIC PN, and up to $\sim$ 130 keV
by analysing RXTE HEXTE data. The morphologies of the EPIC PN (taking into account the coarser timing resolution) and the HEXTE profiles are consistent with that measured with the PCA.

-- The spectrum of the pulsed X-ray emission is of non-thermal origin, exhibiting a power-law shape with photon index $\Gamma$ = 1.06 $\pm$ 0.03, fitting just the high-statistics PCA data, and $\Gamma$ = 1.03 $\pm$ 0.02, fitting over the broader energy
band from $\sim$ 0.5 to $\sim$ 270 keV by including also the EPIC-PN and HEXTE flux values. There is no indication for a black-body component in the soft X-ray spectrum above 0.5 keV.

-- We do not see a spectral difference between the spectra of the two X-ray pulses in the PCA data. Both spectral photon indices are fully consistent with the time averaged value for the total pulsed emission (see Table \ref{spectral_fit}). Note that the relative strengths of P1 and P2 in the X-ray and high-energy $\gamma$-ray windows reverse (see Fig. \ref{xrg_collage}); the spectrum of P2 has to harden significantly with respect to that of P1 between a few hundred keV and 100 MeV.

\section{Discussion and Conclusions}

\label{sect_disc}
\begin{figure}
 \centering
  \includegraphics[width=8.8cm,height=8cm]{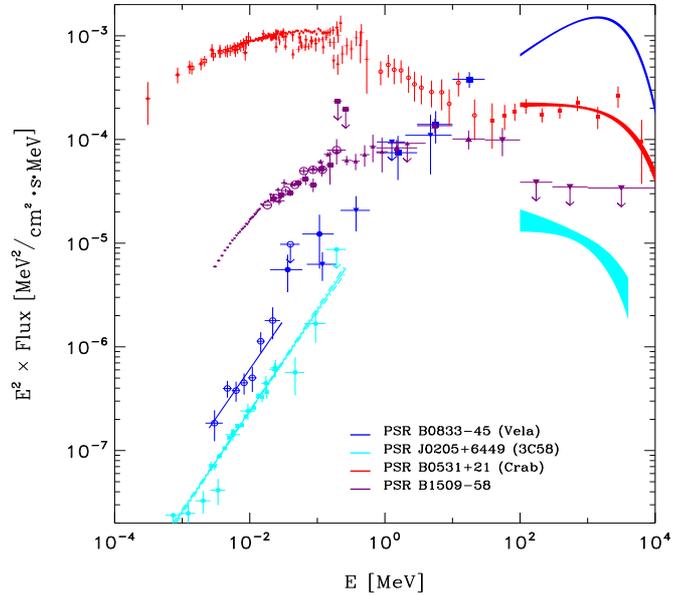}
   \caption{Broad-band total-pulsed photon spectrum of \psrtar\ (aqua colored) compared with the pulsed spectra of PSR B0531+21
           (Crab; red), PSR B0833-45 (Vela; dark blue) and PSR B1509-58 (purple). The (hard) X-ray spectrum (0.56-267.5 keV) of 
           \psrtar\ has been derived in this work, and the best-fit power-law model (index $\sim$ 1.03) has been superposed. 
            The $>100$ MeV spectrum of \psrtar\ is the model fit to the Fermi spectrum from \citet{abdo2009a}.
            Also for Crab and Vela the best-model fits to the recently published Fermi spectra for energies $>100$ MeV are shown
            \citep[see][respectively]{abdo2009b,abdo2009d}. For all spectra the interstellar absorption has been modeled out 
            (only effective below $\sim 5$ keV).}
\vspace{0.0cm}
\label{he_spectrum}
\end{figure}

In the introduction we noted that \psrtar\ is now one of only three young ($<10,000$ year old) pulsars which are detected in the classical X-ray band and at hard X-rays above 20 keV, as well as at high-energy ($>100$ MeV) $\gamma$-rays, the others being the Crab pulsar and PSR B1509-58.  Fig. \ref{he_spectrum} shows that these three young pulsars reach their maximum luminosities below 100 MeV, 
while the ``middle-aged'' Vela pulsar (characteristic age 11.4 kyr) reaches its maximum at GeV energies \citep[for the latest results on the Vela pulsar for energies above 100 MeV, see the Fermi results by][]{abdo2009b}. The latter spectrum is characteristic for older pulsars
reported to be detected above 100 MeV \citep[e.g. see the first Fermi Large Area Telescope catalog of $\gamma$-ray pulsars by][]{abdo2009c}

Comparing in more detail the high-energy spectra of the young pulsars in Figure \ref{he_spectrum}, we notice large differences.
For energies below 10 keV the flux values of \psrtar\ are $\sim$ 4 orders of magnitude below those of the Crab, while around 10 MeV the difference is reduced to about a factor of 10. The X-ray spectrum of \psrtar\ is, thus, very much harder than that of the Crab. 
The total high-energy spectrum of \psrtar\ appears to reach its maximum luminosity at MeV energies, like is the case for PSR B1509-58/PSR J1513-5908.
The spectral break for the latter spectrum between 10 MeV and 100 MeV (see flux values in Figure \ref{he_spectrum}) measured with COMPTEL and EGRET aboard the Compton Gamma-Ray Observatory by \citet{kuiper1999} has been confirmed by \citet{pellizzoni2009}. These authors report for 
PSR B1509-58 a softening of the photon index $\Gamma$ from $\sim$1.7 to $\sim$2.5 going from tens to hundreds of MeV (but do not provide pulsed-flux values).

{\tabcolsep=1.25mm
\begin{table*}[t]
\caption{Characteristics of the three young ($<10$ kyr) X-ray and $\gamma$-ray emitting pulsars in comparison with the middle-aged Vela pulsar 
(PSR B0833-045). The luminosities of the pulsed emission $L$ are calculated as $L = 4\pi d^2Ff_{\Omega}$ with values for the distance $d$ taken from the table, and the beaming fraction $f_\Omega$ set to 1. $F$ represent the pulsed flux.}
\label{pulsar_comparison}
\begin{center}
\begin{tabular}{lcccccccccc}
\hline
Source               &  d   &  P   &  $\tau$ & $L_{sd}$ & $F_x^a$         & $F_{\gamma}^b$ & $L_x^a$   & $\eta_x^a$ & $L_{\gamma}^b$ & $\eta_{\gamma}^b$\\
                     &[kpc] & [ms] &  [ky]   & [erg/s]  & [erg/cm$^2$s] & [erg/cm$^2$s] & [erg/s] &          &  [erg/s]     &                \\
\hline\hline\\
\vspace{-4mm}\\
B0531+21   &  2.0     &  29.7   &  1.3         &4.4E+38 & $(5.68\pm 0.05)$E-09 & $(1.3\pm0.1)$E-09$^{\ }$ & $(2.72\pm0.02)$E+36 & $6.2$E-3 & $(6.1\pm0.3)$E+35$^{\ }$ & $1.4$E-3$^{\ }$\\
B1509-58   &  5.8     & 151.5   &  1.6         &1.7E+37 & $(1.46\pm 0.02)$E-10 & $(5.1\pm2.5)$E-11$^{c}$ & $(5.88\pm0.08)$E+35 & $3.5$E-2 & $(2.1\pm1.0)$E+35$^{c}$ & $1.2$E-2$^{c}$\\
J0205+6449 &  3.2     &  65.7   &  5.4         &2.7E+37 & $(0.36\pm 0.02)$E-11 & $(6.7\pm0.5)$E-11$^{\ }$ & $(4.45\pm0.20)$E+33 & $1.7$E-4 & $(8.2\pm0.6)$E+34$^{\ }$ & $3.0$E-3$^{\ }$\\
B0833-045  &  0.287   &  89.3   & 11.4         &6.9E+36 & $(0.88\pm 0.29)$E-11 & $(7.9\pm0.3)$E-09$^{\ }$ & $(8.67\pm2.85)$E+31 & $1.3$E-5 & $(7.8\pm0.3)$E+34$^{\ }$ & $1.1$E-2$^{\ }$\\
\hline\hline\\
\vspace{-4mm}\\
\multicolumn{11}{l}{$^a$ Luminosities, fluxes and efficiencies labeled with $x$ are evaluated for the 2-100 keV band.}\\ 
\multicolumn{11}{l}{$^b$ Luminosities, fluxes and efficiencies labeled with $\gamma$ are evaluated for the 0.1-10 GeV band.}\\ 
\multicolumn{11}{l}{$^c$ The $\gamma$-ray energy flux of PSR B1509-58 in the 0.1-10 GeV band has been derived from the (total) photon flux values for the 100-300 and }\\
\multicolumn{11}{l}{\ \  300-1000 MeV bands as given in \citet{kuiper1999} assuming a power-law shape with photon index of $2.5$, and should be considered as}\\ 
\multicolumn{11}{l}{\ \  an upper-limit to the pulsed flux of PSR B1509-58 in the 0.1-10 GeV band.}\\
\end{tabular}
\end{center}
\end{table*}
}

Interestingly, the X-ray spectrum above 2 keV of \psrtar\ resembles that of the slightly older Vela pulsar (similar spectral index), 
but the $L_X/L_{\gamma}$ ratio for the pulsed component differs by a factor $\sim$50; higher for \psrtar\ (see Table \ref{pulsar_comparison}, which is introduced below). The $L_X/L_{\gamma}$ ratio of \psrtar\ is in between those of Vela and PSR B1509-58, namely, the $L_X/L_{\gamma}$ ratio for \psrtar\ is a factor $\sim$50 smaller than that for PSR B1509-58. Note, that for the quoted flux ratios it is assumed that the beaming fractions in the X-ray and $\gamma$-ray bands are the same. We know, however, that these are in many cases different. More importantly, in the X-ray spectra below e.g. 2 keV there are no indications for black-body components in the spectra of Crab, PSR B1509-58 and \psrtar. 
In contrast, the (pulsed) Vela spectrum exhibits below 2 keV a black-body peak \citep[not shown in Fig. \ref{he_spectrum}; see e.g.][]{pavlov2001}, which is characteristic for middle-aged and older rotation powered pulsars. Therefore, the spectral properties of \psrtar\ confirm that we are dealing with a young pulsar, and suggest a real age between those of Vela and PSR B1509-58, favouring its characteristic age of 5.4 kyr over that of SN 1181 (828 yr).

Table \ref{pulsar_comparison} lists for the four pulsars discussed above in order of characteristic age ($\tau=P/2\dot{P}$) the spin-down luminosities $L_{sd}$ and fluxes $F$ and luminosities $L$ in the X-ray 2-100 keV and gamma-ray 0.1-10 GeV bands, as well as the corresponding efficiencies to convert spin-down energy into emission in these energy bands. The luminosities are calculated as $L = 4\pi d^2Ff_\Omega$, with the values for the
distance $d$ taken from the table, and the value for $f_\Omega$, which is the beaming fraction, set to 1. At first sight one could argue that there is an evident anticorrelation between characteristic age and X-ray luminosity, independent from differences in the beaming fractions, but this becomes less obvious when we consider the X-ray efficiencies instead of luminosities. In the gamma-ray band there is not any indication for a correlation. The listed gamma-ray efficiencies differ less than a factor $\sim$10, ignoring differences in beaming fraction, while the latter differences can be substantial. 

There are also large differences in the morphologies of the pulse profiles of the three young pulsars. Comparing the pulse profiles detected for Crab and \psrtar\ at X-ray energies and high-energy $\gamma$-rays, then there are also some simularities: both exhibit two pulses with peaks separated $\sim$0.5 and $\sim$0.4 in pulse phase, respectively, and the X-ray and $\gamma$-ray pulses are aligned in phase. However, the pulses in the Crab profile are significantly broader than those of \psrtar\, and emission is also detected between the two Crab pulses. The latter is not the case for the X-ray profile of \psrtar, but, interestingly, emission between the pulses has been detected in the Fermi profile at the $5\sigma$ level.

Furthermore, the Crab main radio pulse is in general phase coincidence with the broad X-ray/$\gamma$-ray pulse. The peak of this main radio pulse lags that of the first X-ray/$\gamma$-ray pulse in phase by only $\sim0.008$ or 280 $\mu$s; see for consistent estimates from INTEGRAL, RXTE and EGRET \citet{kuiper2003}, and from Fermi \citet{abdo2009d}. On the other hand, the Crab radio precursor preceeds the first, main X-ray/$\gamma$-ray pulse in phase by $\sim$0.09, or 3.2 ms, being located around the start of the leading wing of the high-energy pulse. 
In the case of \psrtar, the single radio pulse is also preceeding the first narrow X-ray/$\gamma$-ray pulse in phase by $\sim$ 0.083 or 5.4 ms, and is fully separated in phase, the radio pulse being located just before the onset of the first high-energy pulse (see Fig. \ref{xrg_collage}). This strongly suggests that the analogue of the radio pulse of \psrtar\ is the weak radio precursor of the Crab. Also, that there are no counterparts in the radio profile of \psrtar\ to the two high-energy pulses of \psrtar, contrary to the situation for the Crab. This means that for this young pulsar exhibiting sharp non-thermal high-energy pulses, we do not see evidence for radio emission originating from the same site in the magnetosphere, e.g. in slot gaps \citep[two-pole caustic emission,][]{dyks2003} or outer gaps 
\citep[outer-magnetosphere emission, see][from a region close to the light cylinder]{cheng1986,romani1996,hirotani2006}.  
Possibly, this radio component of \psrtar\ is just too weak to be detectable, but might be revealed in a search for giant radio pulses in the phase intervals of the high-energy pulses. Namely, for a number of young and milli-second radio pulsars phase coincidences between the high-energy pulses and giant radio pulses have been reported. Two examples: the Crab for which the distribution of giant
radio pulses is remarkably similar to the average emission profile of the radio main and interpulse \citep{popov2006} and 
milli-second pulsar PSR B1937+21 for which \citet{cusumano2003} reported the phase coincidence of two sharp high-energy X-ray pulses with two phase intervals exhibiting giant radio pulses, which trail the two normal radio pulses. The latter example might be revealed for \psrtar.

The high-energy pulse profile of PSR B1509-58 differs totally from those of Crab and \psrtar. At hard X-rays and soft $\gamma$-rays below 10 MeV the profile consists of a single structured broad pulse, which can be explained as being composed of two Gaussian pulse profiles separated $\sim$0.14 in phase with different spectra \citep{kuiper1999,cusumano2001}, the second broader pulse peaking at $\sim$0.35, with the main radio pulse at phase 0. Above 10 MeV, the COMPTEL profile between 10 and 30 MeV and the EGRET profile between 30 and 100 MeV suggest the presence of an additional high-energy pulse at phase $\sim$0.85 \citep{kuiper1999}. The latter seems now to be confirmed in the AGILE profile of PSR B1509-58 \citet{pellizzoni2009}, which shows the main pulse for energies above 100 MeV at phase $\sim$0.35, and a second possible pulse at $\sim$0.85. It is now ambiguous what phase difference between high-energy pulses of PSR B1509-58 ($\sim$0.14 or $\sim$0.5) should be considered for comparison with the morphology of pulse profiles of the other young pulsars. 

The {\bf above cited} different models aiming to explain the production of non-thermal high-energy emission in the magnetospheres of rotation-powered pulsars do not address flux variability. Furthermore, there was also no observational evidence for such variability till the magnetar-like outburst of the high-field pulsar PSR J1846-0258 \citep{gavriil2008}, which decayed with an $1/e$-time constant of $\sim$ 55 days.
It was shown by \citet{kuiper2009} that the radiative outburst was triggered by a major spin-up glitch, and that, most interestingly, the shape 
of the X-ray pulse profile did not change during the outburst. For the flux increase by a factor of $\sim$ 2 of the non-thermal emission  from the second pulse of \psrtar\ during a two-week time period, we did not see a variation in pulse shape, either. However, there was no
indication for glitching activity. The significance of the variability is insufficient to draw strong conclusions, but it seems waranted to start searching for such variability  in the emission from the increasing sample of rotation-powered pulsars emitting non-thermal high-energy emission.  

In conclusion, we accurately measured for the young rotation powered pulsar \psrtar\ the morphology of the X-ray light curve and the spectrum over the broad X-ray band $\sim$0.5 - $\sim$270 keV.  The \psrtar\ X-ray spectrum above 2 keV has the same power-law shape ($\Gamma \sim$ 1.03) as the middle-aged Vela pulsar, but the overall high-energy spectral shape, considering also the Fermi $\gamma$-ray spectrum, resembles more the spectrum expected for a younger pulsar, i.e. no evidence for a thermal black-body component, and maximum luminosity at MeV energies and not at GeV energies. 

The morphology of the double-pulse \psrtar\ light curve can be explained in a conventional outer-gap scenario for a rotating dipole in vacuum assuming low-altitude radio emission, similar to the case of the Crab pulsar when taking the Crab precursor radio pulse as the counterpart of the single radio pulse detected for \psrtar. This can be verified in the ``Atlas'' of model $\gamma$-ray light curves simulated by \citet{watters2009}. However, see also the alternative Atlas by \citet{bai2009a}, who point out an inconsistency in the model calculations by \citet{watters2009} and in earlier reports, affecting particularly profile shapes calculated for the two-pole caustic model.

Furthermore, it should be realized that the sharp pulses in the high-energy profile of this young pulsar \psrtar\ do not have radio counterparts like we see for the Crab, and we encourage a search for giant radio pulses in the phase intervals of these high-energy pulses. Recent model calculations by \citet{bai2009b} using a force-free field instead of the vacuum dipole field show that alternative scenarios such as their annular-gap model are required to produce over a wide range of parameters two sharp high-energy pulses as
exhibited by  \psrtar. More extensive 3-D simulations including the physics of the production processes are required for more detailed comparisons with the spectral and timing characteristics.

\begin{acknowledgements}
This research has made use of data obtained from the High Energy Astrophysics 
Science Archive Research Center (HEASARC), provided by NASA's Goddard Space Flight Center.
We have extensively used NASA's Astrophysics Data System (ADS). JOU acknowledges the IAU Travel Grant that
enabled him to visit SRON; and the support and hospitality of SRON Netherlands Insitute for Space Research.
\end{acknowledgements}

\end{document}